**Recurrent frequency-size distribution of characteristic events**

**S.G. Abaimov[1]**

[1]Department of Geology, University of California, Davis, CA, 95616, USA.

Email: sgabaimov@ucdavis.edu

Accepted          Received               In original form

## SUMMARY

Many complex systems, including sand-pile models, slider-block models, and earthquakes, have been discussed whether they obey the principles of self-organized criticality. Behavior of these systems can be investigated from two different points of view: interoccurrent behavior in a region and recurrent behavior at a given point on a fault or at a given fault. The interoccurrent frequency-size statistics are known to be scale-invariant and obey the power-law Gutenberg-Richter distribution. This paper investigates the recurrent frequency-size behavior of characteristic events at a given point on a fault or at a given fault. For this purpose sequences of creep events at a creeping section of the San Andreas fault are investigated. The applicability of the Brownian passage-time, lognormal, and Weibull distributions to the recurrent frequency-size statistics of slip events is tested and the Weibull distribution is found to be a best-fit distribution. To verify this result the behaviors of the numerical slider-block and sand-pile models are investigated and the applicability of the Weibull distribution is confirmed. Exponents $\beta$ of the best-fit Weibull distributions for the observed creep event sequences and for the slider-block model are found to have close values from 1.6 to 2.2 with the corresponding aperiodicities $C_V$ of the applied distribution from 0.47 to 0.64.

## 1 INTRODUCTION

A concept of self-organized criticality (SOC) has been introduced by Bak *et al.* (1988). Many studies have investigated this concept and its connection with critical phenomena in statistical mechanics. Some researchers associate the concept of SOC with systems exactly at a critical point where power-law dependences prevail (Bak *et al.* 1988). Others attribute more general understanding to this concept. Further in this paper a reference to the concept of SOC will mean a system with avalanches that can evolve to a critical state by tuning some external boundary parameters. For example, by saying that a slider-block model obeys the principles of SOC we will imply that the behavior of avalanches in this model tends to a critical state when its stiffness tends to infinity. In other words, we attribute the term 'self-organized criticality' rather to the presence of time-clustering and avalanche behavior than to the space-clustering and critical properties which are introduced in statistical mechanics and do not require a special name. If a reader prefers more rigorous definition restricted to the original meaning introduced by Bak *et al.* (1988) it will be relevant to read 'near-critical phenomena of dynamical systems with avalanches' instead of 'self-organized criticality'.

Many complex systems have been discussed whether they obey the principles of SOC. These systems in general exhibit two different types of statistical behavior. In this paper we will follow the terminology introduced by Abaimov et al. (2007b). The term '*interoccurrent*' we will use for statistics of earthquakes on all faults in a region. The term '*recurrent*' will be used for event sequences of a single fault or fault segment. These events will be referred to as *characteristic* events if they have more or less the same rupture area equivalent to the area of the fault or fault segment (if they rupture more or less the same set of asperities). The discussion of correspondence between characteristic events and sets of asperities can be found in, *e.g.* (Okada *et al.* 2003; Park and Mori 2007).

The *interoccurrent frequency-size* statistics have been investigated by many authors and have been shown to obey the power-law Gutenberg-Richter distribution. This type of interoccurrent scale-invariant behavior has been found for earthquakes (Gutenberg and Richter 1954), simulations of the slider-block model (Carlson and Langer 1989), and for the sand-pile model (Bak et al. 1988). All these systems have been discussed in the literature whether they obey (under some conditions) the principles of SOC.

To estimate hazard risks for a region it is necessary to know the statistical properties of earthquake occurrence. The interoccurrent statistics play an important role in these assessments. Therefore many studies have been devoted to the investigation of the interoccurrent behavior of earthquakes. However, the knowledge of only interoccurrent properties is not sufficient. It is also necessary to know the statistical properties of another type of behavior - the *recurrent* behavior at a given point on a fault or at a given fault. Unfortunately, about this specific type of behavior we have little information. The reason is that this type of behavior is much more difficult to be investigated. For the interoccurrent frequency-size statistics of a region it is only necessary to count magnitudes of earthquakes occurred in this region. For the recurrent behavior it is required to associate these earthquakes each to a specific fault or fault segment. Actually, the question what statistics correspond to the recurrent behavior of earthquakes has not been answered yet. In the literature primarily the recurrent time-interval statistics have been investigated (Abaimov *et al.* 2007a; Abaimov *et al.* 2007b; Hagiwara 1974; Jackson *et al.* 1995; Matthews *et al.* 2002; Molchan 1990, 1991; Nishenko and Buland 1987; Ogata 1999; Rikitake 1976, 1982, 1991; Utsu 1984; Working Group on California Earthquake Probabilities 1988, 1990, 2003). Only a few attempts have been made to investigate the *recurrent frequency-size* statistics (e.g., Abaimov *et al.* 2007a; Bakun *et al.* 2005).

This paper overcomes this lack. As some possible alternatives for the recurrent frequency-size behavior we will consider the Brownian passage-time, lognormal and Weibull distributions. In Section 2 we briefly describe these distributions. The tests of goodness-of-fit that we use are described in Section 3. These include the Kolmogorov–Smirnov test and the root mean squared error test.

Ideally, observed sequences of earthquakes on a fault should be used to establish the applicable statistical distribution. However, the numbers of events in observed earthquake recurrent sequences are not sufficient to establish definitively the validity of a particular distribution (Savage 1994). To illustrate this in Section 4 we will consider the sequence of earthquakes on the Parkfield segment of the San Andreas fault.

In Section 5 of this paper we study sequences with a large number of recurrent events. For this purpose we investigate a creeping section of the San Andreas fault. Creep events on the central section of the San Andreas fault have been studied extensively. The creep measurements have been carried out since the 1960s by U.S. Geological Survey and show both steady-state creep and well defined slip events. We consider the recurrent statistics of slip events that are superimposed on the steady-state creep. For the four sequences we test the applicability of the Brownian passage-time, lognormal, and Weibull distributions to the recurrent frequency-size statistics of slip events. In this case, we have enough events to convincingly differentiate between alternative proposed distributions. In each case we compare the data (sample distribution) with the three distributions and provide tests of goodness-of-fit.

Although the creep records provide enough (up to 100) events in a sequence to differentiate between the proposed distributions, the applicability of these statistics to earthquakes could be questionable. Therefore, it would be reasonable to test our results with the aid of numerical simulations. A slider-block is often used to study earthquake behavior (see e.g. Abaimov *et al.* 2007b; Abaimov *et al.* 2008; Carlson and Langer 1989). Although the applicability of this model to earthquakes is often questionable too, this would give us an independent verification of our results. Therefore in Section 6 we investigate the recurrent behavior of the slider-block model.

A sand-pile model was a basic model for developing SOC ideas (Bak et al. 1988). Therefore it would be curious to see what recurrent behavior this model has. We investigate this question in Section 7.

## 2 APPLICABLE DISTRIBUTIONS

Three widely used statistical distributions in geophysics are the Brownian passage-time, lognormal, and Weibull. We will compare each of them with our data (sample distribution).

## 2.1 BROWNIAN PASSAGE-TIME DISTRIBUTION

The probability density function (pdf) of amplitudes $S$ for the Brownian passage-time distribution is given by (Chhikara and Folks 1989)

$$p(S) = \left( \frac{\mu}{2\pi C_V^2 S^3} \right)^{\frac{1}{2}} \exp\left[ -\frac{(S-\mu)^2}{2C_V^2 \mu S} \right] \tag{1}$$

where $\mu$ is the mean, $\sigma$ is the standard deviation, and $C_V = \frac{\sigma}{\mu}$ is the aperiodicity (coefficient of variation) of the distribution. The corresponding cumulative distribution function (cdf) can be obtained analytically (Matthews et al. 2002) but the expression is lengthy and is not given here explicitly.

## 2.2 LOGNORMAL DISTRIBUTION

The lognormal is one of the most widely used statistical distributions in a wide variety of fields. The pdf of amplitudes $S$ for the lognormal distribution is given by (Patel et al. 1976)

$$p(S) = \frac{1}{(2\pi)^{1/2} \sigma_y S} \exp\left[-\frac{(\ln S - \bar{y})^2}{2\sigma_y^2}\right] \qquad (2)$$

The lognormal distribution can be transformed into the normal distribution by making the substitution $y = \ln S$; $\bar{y}$ and $\sigma_y$ are the mean and standard deviation of this equivalent normal distribution. The mean $\mu$, standard deviation $\sigma$, and aperiodicity (coefficient of variation) $C_V$ for the lognormal distribution are given by

$$\mu = \exp\left[\bar{y} + \frac{\sigma_y^2}{2}\right], \ \sigma = \mu\sqrt{e^{\sigma_y^2} - 1}, \text{ and } C_V = \frac{\sigma}{\mu} = \sqrt{e^{\sigma_y^2} - 1}. \qquad (3)$$

The corresponding cdf $P(S)$ (fraction of amplitudes that are smaller than $S$) for the lognormal distribution is

$$P(S) = \frac{1}{2}\left(1 + erf\left[\frac{\ln S - \bar{y}}{\sqrt{2}\sigma_y}\right]\right) \qquad (4)$$

where $erf(x) = \frac{2}{\sqrt{\pi}} \int_0^x e^{-y^2} dy$ is the error function.

## 2.3 WEIBULL DISTRIBUTION

The Weibull distribution is often used in engineering applications (Meeker and Escobar 1991; Weibull 1951). The pdf for the Weibull distribution is given by (Patel et al. 1976)

$$p(S) = \frac{\beta}{\tau}\left(\frac{S}{\tau}\right)^{\beta-1} \exp\left[-\left(\frac{S}{\tau}\right)^{\beta}\right] \qquad (5)$$

where $\beta$ and $\tau$ are fitting parameters. The mean $\mu$ and the aperiodicity (coefficient of variation) $C_V$ of the Weibull distribution are given by

$$\mu = \tau\, \Gamma\left(1+\frac{1}{\beta}\right) \qquad (6)$$

$$C_V = \left\{ \frac{\Gamma\left(1+\frac{2}{\beta}\right)}{\left[\Gamma\left(1+\frac{1}{\beta}\right)\right]^2} - 1 \right\}^{\frac{1}{2}} \qquad (7)$$

where $\Gamma(x)$ is the gamma function of $x$. The cdf for the Weibull distribution is given by

$$P(S) = 1 - \exp\left[-\left(\frac{S}{\tau}\right)^{\beta}\right] \tag{8}$$

If $\beta = 1$ the Weibull distribution becomes the exponential distribution with $\sigma = \mu$ and $C_V = 1$. In the limit $\beta \to +\infty$ the Weibull distribution becomes exactly repetitive (a δ-function) with $\sigma = C_V = 0$. In the range $0 < \beta < 1$ the Weibull distribution is often referred to as the stretched exponential distribution.

## 3 TESTS OF GOODNESS-OF-FIT

In order to determine whether a specific distribution is preferred, it is necessary to utilize tests of goodness-of-fit. Many such tests are available (Press et al. 1995). In this paper we quantify the goodness-of-fit of distributions using two tests.

## 3.1 KOLMOGOROV-SMIRNOV TEST

The first test we use is the Kolmogorov-Smirnov test (Press et al. 1995). To use this test the maximum absolute difference $D_{KS}$ between the cdf of the sample distribution (actual data) $y_i$ and the fitted distribution $\bar{y}_i$ is determined:

$$D_{KS} = \max|y_i - \bar{y}_i| \tag{9}$$

Then the significance level probability of the goodness-of-fit (the probability that the applied distribution is relevant) is given by

$$Q_{KS}(\lambda) = 2\sum_{i=1}^{+\infty}(-1)^{i-1}e^{-2i^2\lambda^2} \tag{10}$$

where

$$\lambda = D_{KS}\left(\sqrt{n} + 0.12 + \frac{0.11}{\sqrt{n}}\right), \tag{11}$$

and $n$ is the number of data points. The preferred distribution has the smallest value of $D_{KS}$ and the largest value of $Q_{KS}$.

## 3.2 ROOT MEAN SQUARED ERROR TEST

The second test of goodness-of-fit we use is the root mean squared error (RMSE). As it is suggested by the test name, it is the square root of the sum of squares of errors divided by the difference between the number of data points and the number of fitting parameters

$$RMSE = \sqrt{\frac{\sum_{i=1}^{n}(y_i - \hat{y}_i)^2}{n-k}} \tag{12}$$

where $y_i$ are the sample distribution (actual data), $\bar{y}_i$ are predicted fit values, $n$ is the number of data point, and $k$ is the number of fitting parameters ($k = 2$ for our three distributions). This test is also known as the fit standard error and the standard error of the regression. The preferred distribution has the smallest RMSE value.

## 4 PARKFIELD SEQUENCE

Ideally, recurrent sequences of earthquakes should be used to establish a preferred statistical distribution. Unfortunately, the number of earthquakes available through historical records is too small for adequate statistical testing.

As an example we consider the sequence of seven characteristic earthquakes that occurred on the Parkfield (California) section of the San Andreas fault between 1857 and 2004 (Bakun et al. 2005). The slip rate is quite high ($\approx 30$ mm/year) and the earthquake magnitudes are relatively small ($m \approx 6.0$), thus the recurrent times are short ($\approx 25$ years). Also, this fault is subject to a near constant tectonic drive due to the relative motion between the Pacific and North American plates. Slip on the Parkfield section of the San Andreas fault occurred in 1881, 1901, 1922, 1934, 1966, and 2004 with magnitudes from 6.0 to 6.05 by the instrumental estimate and from 5.9 to 6.1 by the estimate from MMI for an epicenter location on the 2004 rupture (Bakun *et al.* 2005).

As the size of an event we can use here the seismic moment or energy of this event. However, the small number of registered earthquakes makes the application of statistical estimations impossible (Savage 1994). And this problem is relevant not only for the Parkfield sequence. Other earthquake sequences also have short statistics (e.g., Okada *et al.* 2003; Park and Mori 2007). Although it is difficult to register small magnitude earthquakes themselves, there is another problem when one tries to reconstruct the recurrent statistics on a particular fault. For large magnitude earthquakes, like the Parkfield sequence, it is possible to associate the events with a particular fault. But the sequences are short. In contrast, for small magnitude earthquakes it is generally impossible to reconstruct the recurrent statistics due to the difficulty of associating the earthquake waveform with the rupture of a particular fault. Even close locations and waveforms could belong to different faults for small magnitude earthquakes. And, *vise versa*, different waveforms could be generated by the same fault. Therefore it is impossible to solve the problem using only registered earthquake sequences.

The sequence of Parkfield earthquakes gives us only a cue that the actual recurrent frequency-size distribution of characteristic events is much more repetitive (has lower aperiodicity) than the interoccurrent power-law Gutenberg-Richter distribution. Actually, today an exactly repetitive distribution ($\delta$-function, all magnitudes are equal) is used in the probabilistic seismic hazard assessments such as the most recent for the San Francisco Bay region (Working Group on California Earthquake Probabilities 2003). Is the recurrent frequency-size distribution of characteristic events indeed exactly repetitive? Do earthquake magnitudes have no variability? Is there an actual statistical distribution that should be used instead of the $\delta$-function? This paper answers these questions.

## 5 SLIP EVENTS ON A CREEPING SECTION OF THE SAN ANDREAS FAULT

We now consider the recurrent statistics of slip events on the creeping section of the San Andreas fault, California. To do this we utilize records of two creepmeters on the San Andreas fault (Schulz 1989). One of these is located near the Cienega Winery, 16.9 km southeast of San Juan Bautista (station "cwn1", latitude 36° 45.0', longitude 121° 23.1'). The creep measurements have been carried out since June 1972 by the U.S. Geological Survey, and show that the average long-term creep rate is about 11.5 mm/year. The second creepmeter is located near Harris Ranch, 12.8 km southeast of San Juan Bautista, and 4.1 km northwest from cwn1 (station "xhr2", latitude 36° 46.3', longitude 121° 25.3'). The creep measurements have been carried out since April 1985 by the U.S. Geological Survey, and show that the average long-term creep rate is in the range from 6 mm/year to 9 mm/year. The recorded data for both creepmeters can be downloaded from the U.S. Geological Survey web site (Langbein 2004).

For both creepmeters the data contain both daily and 10 minute telemetry records. Although the daily records are longer and give longer sequences, the 10 minute data also are used independently because they provide more accurate slip amplitude resolution.

Because the creepmeter piers are installed at shallow depths (2-3 meters), the piers could tilt in response to rainfall. Therefore the cwn1 10 minute record has been investigated together with the records of the nearest rainfall station (station "CHT", Pajaro river basin at Chittenden, latitude 36.9020°, longitude 121.6050°, operator U.S. Geological Survey and California Department of Water resources (California Department of Water Resources 2006)). No significant influence on the procedure of event determination has been found; properties of no one event have been corrected. Therefore the influence of rainfalls for other records has not been examined.

For earthquakes the duration of an event is of the order of seconds to minutes and can be neglected in comparison with the preceding period of slow stress accumulation. In contrast, for creep records, the duration of an event can be of the order of the interval between events. Therefore special techniques are required to distinguish one event from the next. To do this we use the criterion of stationarity. If, after a well defined jump, the slip rate returns to a stationary creeping state before the next jump, these two jumps are treated as separate events. Otherwise, if one jump triggers another one in a transient, non-stationary process, these jumps are considered to be a single event.

Each creep record provides a unique opportunity to determine the complete sequence of events taking place at a given creepmeter location. In contrast to earthquakes, the rate of occurrence for creep events is much higher. Also each observed sequence of creep events provides the complete record of all events occurred at a given location. This gives an opportunity to associate events not only with the given fault but even with the given point of this fault. Therefore the reconstruction of event sequences from creep records gives both the longest possible sequences (up to 100 events) and the most accurate determination of recurrent events.

To filter the telemetry noise, threshold levels for slip amplitude are used: 0.077 mm for 10 minute xhr2 telemetry, 0.078 mm for 10 minute cwn1 telemetry, 0.3 mm for xhr2 daily telemetry, and 0.31 mm for cwn1 daily telemetry.

As the size of an event we will use the slip amplitude of this event. The cumulative recurrent frequency-amplitude distributions of slip events for the 10 minute xhr2 and cwn1 records are given in Figs. 1(a) and 1(b) *on* *linear axes*. For each value of slip amplitude $S_0$ the $N(S \geq S_0)$ is the number of events with amplitudes $S$ greater than $S_0$. The data show a smooth dependence except for small amplitudes. The anomaly of small amplitudes could be associated with the presence of non-characteristic events. Therefore further to obtain sequences of characteristic events we will use amplitude threshold 0.3 mm and will discard small amplitude events below this threshold. This approach appears to be reasonable. Indeed, visual comparison of records for two different locations xhr2 and cwn1 separated by 4.1 km suggests that slip amplitudes below 0.3 mm have in general extensions less than the distance between these two locations while slip amplitudes above 0.3 mm can be traced from one record to another.

The cumulative recurrent frequency-amplitude distributions of slip events for the xhr2 and cwn1 daily records are given in Figs. 1(c) and 1(d) *on linear axes*. Here the low resolution of the telemetry data does not allow us to distinguish small amplitude events, thus the anomaly of small amplitudes is absent in these Figures. Therefore we assume that these sequences already contain only characteristic events and the telemetry resolution acts here as the amplitude threshold 0.3 mm.

The frequency-amplitude distributions given in Figs. 1(a), 1(b), 1(c), and 1(d) clearly are not the fractal Gutenberg-Richter statistics associated with the frequency-magnitude distribution of earthquakes in a region. This is not surprising since Gutenberg-Richter statistics are associated with earthquakes that occur on many faults. For the recurrent statistics of characteristic events more repetitive (with lower aperiodicity) distributions should be considered.

Frequency-size statistics in Figs. 1(a), 1(b), 1(c), and 1(d) have been constructed in the traditional style of the Gutenberg-Richter distribution (except only linear-linear scale for both axes): The cumulative distribution function (cdf) integrates events from large to small amplitudes and is not normalized. However, we could construct the same plots as it would be done by statisticians. For this purpose we first introduce amplitude thresholds 0.3 mm for both 10 minute records. This excludes non-characteristic events from these sequences. For the xhr2 and cwn1 daily sequences the introduction of this threshold is not necessary because the anomaly of small amplitudes is already beyond the limit of the telemetry resolution 0.3 mm. Finally, we integrate events from small to large amplitudes and divide the result by the total number of events.

The cumulative distribution of 51 recurrent slip amplitudes $P(S)$ for the xhr2 10 minute event sequence is given as a function of the slip amplitude $S$ in Fig. 2(a). The cumulative distribution of 45 recurrent slip amplitudes $P(S)$ for the cwn1 10 minute event sequence is given as a function of the slip amplitude $S$ in Fig. 2(c). The means and coefficients of variation of these recurrent slip amplitudes are given in Table 1. Also included in Figs. 2(a) and 2(c) are the best-fits (maximum likelihood fits) of the Brownian passage-time, lognormal, and Weibull distributions. Both the parameters of these fits and the goodness-of-fit estimators are given in Table 1.

In Figs. 2(b) and 2(d) the recurrent statistics for the xhr2 and cwn1 10 minute event sequences are plotted in the form $-\ln(1-P(S))$ versus $S$ in $\log_{10}$-$\log_{10}$ axes. In this form the Weibull distribution is a straight-line fit with slope $\beta$ so that this is known as a Weibull plot. Also included in these Figures are the previous best-fits of the Brownian passage-time, lognormal, and Weibull distributions with RMSE in the $\log_{10}$-$\log_{10}$ axes given in Table 1.

The cumulative distribution of 76 recurrent slip amplitudes $P(S)$ for the xhr2 daily event sequence is given as a function of the slip amplitude $S$ in Fig. 3(a). The cumulative distribution of 104 recurrent slip amplitudes $P(S)$ for the cwn1 daily event sequence is given as a function of the slip amplitude $S$ in Fig. 3(c). The means and coefficients of variation of these recurrent slip amplitudes are given in Table 1. Also included in Figs. 3(a) and 3(c) are the best-fits (maximum likelihood) of the Brownian passage-time, lognormal, and Weibull distributions. Both the parameters of these fits and the goodness-of-fit estimators are given in Table 1.

Figs. 3(b) and 3(d) present the Weibull plots corresponding to Figs. 3(a) and 3(c) respectively. Also included in these Figures are the corresponding best-fits of the Brownian passage-time, lognormal, and Weibull distributions with the RMSE in $\log_{10}$-$\log_{10}$ axes given in Table 1.

The estimators of goodness-of-fits given above demonstrate convincingly that the fits of the Weibull distribution for all four sequences are much better than the lognormal or Brownian passage-time distributions. In particular, the values of $D_{KS}$ and RMSE for both cdf and Weibull plots are significantly smaller and the values of $Q_{KS}$ are significantly larger for the fits of Weibull distribution than the corresponding values for the fits of other distributions. Also, direct visual tests indicate the strong tendency of the sample distributions to be linear on Weibull plots, *i.e.*, the intrinsic property of the Weibull distribution.

## 3 SLIDER-BLOCK MODEL

In this section we consider the behavior of a slider-block model in order to study the statistics of event sizes. We utilize a variation of the linear slider-block model which Carlson and Langer (1989) used to illustrate the self-organization of such models. We consider a linear chain of 500 slider blocks of mass $m$ pulled over a surface at a constant velocity $V_L$ by a loader plate as illustrated in Fig. 4. Each block is connected to the loader plate by a spring with spring constant $k_L$. Adjacent blocks are connected to each other by springs with spring constant $k_C$. Boundary conditions are assumed to be periodic: the last block is connected to the first one.

The blocks interact with the surface through friction. In this paper we prescribe a static-dynamic friction law. The static stability of each slider-block is given by

$$k_L y_i + k_C \left( 2 y_i - y_{i-1} - y_{i+1} \right) < F_{Si} \qquad (13)$$

where $F_{Si}$ is the maximum static friction force on block $i$ holding it motionless, and $y_i$ is the position of block $i$ relative to the loader plate.

During strain accumulation due to loader plate motion all blocks are motionless relative to the surface and have the same increase of their coordinates relative to the loader plate

$$\frac{dy_i}{dt} = V_L \qquad (14)$$

When the cumulative force of the springs connecting to block $i$ exceeds the maximum static friction $F_{Si}$, the block begins to slide. We include inertia, and the dynamic slip of block $i$ is controlled by the equation

$$m\frac{d^2 y_i}{dt^2} + k_L y_i + k_C(2y_i - y_{i-1} - y_{i+1}) = F_{Di}$$

(15)

where $F_{Di}$ is the dynamic (sliding) frictional force on block $i$. The loader plate velocity is assumed to be much smaller than the slip velocity, requiring

$$V_L << \frac{F_S^{ref}}{\sqrt{k_L m}}$$

(16)

so the movement of the loader plate is neglected during a slip event. The sliding of one block can trigger the instability of the other blocks forming a many block event. When the velocity of a block is zero it sticks with zero velocity.

It is convenient to introduce the non-dimensional variables and parameters

$$\tau_f = t\sqrt{\frac{k_L}{m}}, \quad \tau_s = \frac{tk_L V_L}{F_S^{ref}}, \quad Y_i = \frac{k_L y_i}{F_S^{ref}}, \quad \phi = \frac{F_{Si}}{F_{Di}}, \quad \alpha = \frac{k_C}{k_L}, \quad \beta_i = \frac{F_{Si}}{F_S^{ref}}$$

(17)

The ratio of static to dynamic friction $\phi$ is assumed to be the same for all blocks but the values themselves $\beta_i$ vary from block to block with $F_S^{ref}$ as a reference value of the static frictional force ($F_S^{ref}$ is the minimum value of all $F_{Si}$). Stress accumulation occurs during the slow time $\tau_s$ when all blocks are stable, and slip of blocks occurs during the fast time $\tau_f$ when the loader plate is assumed to be approximately motionless.

In terms of these non-dimensional variables the static stability condition (13) becomes

$$Y_i + \alpha(2Y_i - Y_{i-1} - Y_{i+1}) < \beta_i \tag{18}$$

the strain accumulation Eq. (14) becomes

$$\frac{dY_i}{d\tau_S} = 1 \tag{19}$$

and the dynamic slip Eq. (15) becomes

$$\frac{d^2Y_i}{d\tau_f^2} + Y_i + \alpha(2Y_i - Y_{i-1} - Y_{i+1}) = \frac{\beta_i}{\phi} \tag{20}$$

Before obtaining solutions, it is necessary to prescribe the parameters $\phi$, $\alpha$, and $\beta_i$. The parameter $\alpha$ is a tuning parameter and is the stiffness of the system. We consider $\alpha = 1000$ which corresponds to a very stiff model. We use this high value for the stiffness because the stiff slider-block model obeys the principles of SOC. The ratio $\phi$ of static friction to dynamic friction is taken to be the same for all blocks $\phi = 1.5$, while the values of frictional parameters $\beta_i$ are assigned to blocks by uniform random distribution from the range $1 < \beta_i < 3.5$. This random variability in the system is a "noise" required to thermolize the system and generate event variability.

The loader plate springs of all blocks extend according to Eq. (19) until a block becomes unstable from Eq. (18). The dynamic slip of that block is calculated using the Runge-Kutta numerical method to obtain a solution of Eq. (20). A coupled 4th-order iterational scheme is used, and all equations are solved simultaneously (the Runge-Kutta coefficients of neighboring blocks participate in the generation of the next order Runge-Kutta coefficient for the given block). The dynamic slip of one block may trigger the slip of other blocks and the slip of all blocks is followed until they all become stable. Then the procedure repeats.

For the stiff system with $\alpha = 1000$ the system-wide (500 block) events dominate. If we assume that each block represents an asperity and the whole model represents a fault or fault segment it is reasonable to conclude that the system-wide events correspond to characteristic events. First, we will consider the recurrent statistics at a given point on a fault. In the case of the slider-block model this corresponds to a given block of the model. We consider statistics for the 'strongest', 'weakest', and 'medium' blocks. As a strongest block we choose the block with the highest coefficient of friction, *i.e.,* a block with the highest $\beta_i$. As a weakest block we choose the block with the lowest coefficient of friction. And as a medium block we choose the block with the friction coefficient which is close to the friction coefficient averaged over the model. As the size of an event we choose the total slip amplitude of the given block during this event.

First we construct the cumulative recurrent statistics in the Gutenberg-Richter style. The frequency-amplitude distributions of recurrent slip amplitudes $P(S)$ for event sequences of the strongest, weakest and medium blocks are given as a function of block's slip amplitude $S$ in Figs. 5(a), 5(b), and 5(c) respectively. These Figures are visually similar to Figs. 1(a), 1(b), 1(c), and 1(d). We see here the same anomaly due to the presence of non-characteristic (non-system-wide) events. However, now for the case of numerical simulations we have an opportunity to separate characteristic events directly as system-wide events (without the introduction of the amplitude threshold). Similar to the case of creep events, we remove non-characteristic events from statistics, integrate cdf from small to large amplitudes, and normalize the statistics dividing it by the total number of events.

The cumulative distribution of recurrent slip amplitudes $P(S)$ for the event sequences of the strongest, weakest, and medium blocks are given as functions of the slip amplitude $S$ in Figs. 6(a), 6(c), and 6(e). The means and coefficients of variation of these sequences are given in Table 2. Also included in Figs. 6(a), 6(c), and 6(e) are the best-fits (maximum likelihood) of the Brownian passage-time, lognormal, and Weibull distributions. Both the parameters of these fits and the goodness-of-fit estimators are given in Table 2.

Figs. 6(b), 6(d), and 6(f) present the Weibull plots corresponding to Figs. 6(a), 6(c), and 6(e) respectively. Also included in these Figures are the corresponding best-fits of the Brownian passage-time, lognormal, and Weibull distributions with the RMSE in $\log_{10}$-$\log_{10}$ axes given in Table 2.

The sequences investigated above have been the sequences *at a given point of a fault*. But for the slider-block model we can obtain sequences *at a given fault*. We consider again the system-wide events as the characteristic events. Then we can consider energy dissipated by the whole model (by all blocks) during an event as the size of this event. Indeed, the energy dissipated by all blocks during an event is already not associated with the slip amplitude at a given point of the model but is associated with the slip amplitude averaged over the model. The frequency-size statistics could be constructed both for energies and slip amplitudes – there are no scientific objections to do it anyway. Therefore as the size of an event we can use the energy of this event as well as the slip amplitude.

The cumulative distribution of 715 recurrent energies $P(E)$ for the sequence of system-wide events is given as a function of the dissipated energy $E$ in Fig. 7(a). The mean and coefficient of variation of these recurrent energies are given in Table 3. Also included in Fig. 7(a) are the best-fits (maximum likelihood) of the Brownian passage-time, lognormal, and Weibull distributions. Both the parameters of these fits and the goodness-of-fit estimators are given in Table 3.

Fig. 7(b) presents the Weibull plot corresponding to Fig. 7(a). Also included in this Figure are the corresponding best-fits of the Brownian passage-time, lognormal, and Weibull distributions with the RMSE in $\log_{10}$-$\log_{10}$ axes given in Table 3.

The estimators of goodness-of-fits given above demonstrate convincingly that the fits of the Weibull distribution for all four sequences are much better than the lognormal or Brownian passage-time distributions. In particular, the values of $D_{KS}$ and RMSE for both cdf and Weibull plots are significantly smaller and the values of $Q_{KS}$ are significantly larger for the fits of Weibull distribution than the corresponding values for the fits of other distributions. Also, direct visual tests indicate the strong tendency of the sample distributions to be linear on the Weibull plots, *i.e.*, the intrinsic property of the Weibull distribution.

The frequency-size statistics could be constructed both for energies and slip amplitudes – there are no scientific objections to do it anyway. However, the fact that both statistics have the same behavior (linear on Weibull plot) means that the applied distribution should be invariant relatively to the power-law transformations. Indeed, the energy dissipated by all blocks during an event is associated with the slip amplitude averaged over the model. The seismic moment of a characteristic event is expected to have a power-law dependence on the averaged slip along the fault. So, the logarithm of the energy dissipated during this event by all blocks should be proportional to the logarithm of the averaged slip amplitude. Therefore, if the cumulative distribution of event energies were linear on the Weibull plot then the cumulative distribution of slip amplitudes would be also linear on the Weibull plot. In other words, the functional dependence of the applied distribution should be invariant relatively to the power-law transformations. And the Weibull distribution has this invariant property. This gives an additional verification that the Weibull distribution is the relevant distribution.

So, we have shown that the Weibull distribution is a preferred distribution both for the creep event sequences and for the sequences of characteristic events in the slider-block model. Although the applicability of the behavior of these two systems to earthquakes could be questionable, two independent approaches show converging results.

Another noteworthy fact here is that exponents $\beta$ of the best-fit Weibull distributions for all investigated sequences have similar values from 1.6 to 2.2. In accordance with Eq. (7) the aperiodicities of the best-fit Weibull distributions (and therefore the coefficients of variation $C_V$ of the sample distributions) have close values $C_V = 0.47 \div 0.64$ for both the creep events and the slider-block model.

The same is valid and for the recurrent time-interval statistics. The recurrent time-interval behavior of the characteristic creep events was investigated by Abaimov $et$ $al.$ (2007a). The Weibull distribution was found to be the preferred distribution for the recurrent time-interval statistics and the exponents of the best-fit Weibull distributions had close values from 2.2 to 2.7. This corresponds to the aperiodicity of the applied distribution $C_V = 0.40 \div 0.48$. The recurrent time-interval behavior of the system-wide events of the stiff slider-block model was investigated by Abaimov $et$ $al.$ (2007b). The Weibull distribution was also found to be the preferred distribution for the recurrent time-interval statistics and the exponent of the best-fit Weibull distribution equaled 2.6. This corresponds to the aperiodicity of the applied distribution $C_V = 0.41$ which is close to the case of the creep events $C_V = 0.40 \div 0.48$. Therefore, the creep events and the slider-block model exhibit closely related recurrent behavior of both frequency-size and time-interval statistics. Two independent systems whose applicability to earthquakes is often discussed in the literature have very closely related behavior. This gives us a hope that the stiff slider-block model, which obeys the principles of SOC, could represent the behavior of actual sequences of characteristic earthquakes at a given fault.

## 7 SAND-PILE MODEL

The stiff slider-block model obeys the principles of SOC. Applicability of this concept to earthquakes has been discussed in the literature. Therefore it would be interesting to take a look at the recurrent frequency-size behavior of a sand-pile model as a classical representative of SOC (Bak *et al.* 1988). We utilize the simplest variation of 2-dimensional sand-pile model. A square grid of 100 by 100 sites is strewn by sand grains. When a site accumulates four or more grains it becomes unstable and redistributes four grains to its four neighbors. Instability of one site can trigger instability of others forming a complex avalanche in the system. The sand redistribution during an avalanche is assumed to be much faster than the rate of the stable sand accumulation during strewing (similarly, velocity of earthquake propagation is much faster than tectonic rate of stress accumulation). Therefore the slow sand strewing is neglected during avalanches. Boundaries of the model are assumed to be free so the sand can leave the model at boundaries when it is redistributed by a boundary site.

To construct the recurrent statistics at a given point of the model we choose site in the middle of the lattice. All avalanches with participation of this site will be counted as events at this site of the model. To separate characteristic events we need to construct a criterion similar to the system-wide criterion for the slider-block. For the sand-pile model we choose a percolation criterion as a criterion for an event to be characteristic. *I.e.,* an event will be considered as characteristic if it percolates the lattice and connects all four boundaries (up-right-down-left percolation). As the size of an event we choose the number of different sites participating in this avalanche. Each site can repeatedly loose its stability during this event but will be counted only once in the size of the event.

The cumulative distribution of 2013 recurrent events $P(S)$ is given as a function of the event size $S$ in Fig. 8(a). The mean and coefficient of variation of this statistics are given in Table 4. Also included in Fig. 8(a) are the best-fits (maximum likelihood) of the Brownian passage-time, lognormal, and Weibull distributions. Both the parameters of these fits and the goodness-of-fit estimators are given in Table 4.

Fig. 8(b) presents the Weibull plot corresponding to Fig. 8(a). Also included in this Figure are the corresponding best-fits of the Brownian passage-time, lognormal, and Weibull distributions with the RMSE in $\log_{10}$-$\log_{10}$ axes given in Table 4.

Again, the Weibull distribution is the best-fit distribution. However, now its exponent $\beta$ has much higher value $\beta = 8.09$ which corresponds to the aperiodicity $C_V = 0.15$. This is probably the effect of the percolation criterion as a criterion for an event to be characteristic. Although for this case the concept of self-organized criticality preserves the same functional (Weibull) dependence of the recurrent frequency-size statistics, the sand-pile model has another symmetry and belongs to another universality class with different anomalous dimensions. This question requires further investigation.

## 8 CONCLUSIONS

Recurrent frequency-size distribution plays an important role in the earthquake hazard assessments. However, observed sequences of characteristic earthquakes at a given point on a fault or at a given fault are not able to differentiate among applied statistical distributions. To overcome this difficulty this paper investigates the sequences of creep events at the creeping section of the San Andreas fault. For four sequences the Weibull distribution is shown to be the best-fit distribution. Applicability of this distribution is confirmed by the goodness-of-fit estimators. Direct visual tests also support the applicability of the Weibull distribution because the sample distribution has the tendency to be linear on the Weibull plot.

However, the applicability of slip event statistics to earthquakes could be questionable and an independent verification is required. Therefore two numerical models have been investigated: the stiff slider-block model and the original sand-pile model. Both models support the applicability of the Weibull distributions to the recurrent frequency-size statistics.

Another noteworthy fact is that both the creep event sequences and the sequences of system-wide events of the slider-block model have close values of coefficients of variation $C_V = 0.47 \div 0.64$. The same tendency was also found by Abaimov *et al.* (2007a; 2007b) for the recurrent time-interval behavior of these two systems. The fact that two independent systems whose applicability to earthquakes is often discussed in the literature have closely related recurrent behavior supports the point of view that these two systems actually represent the recurrent earthquake behavior at a given fault or fault segment.

**TABLES**

Table 1. Creep event sequences: The recurrent slip amplitude statistics of creep event sequences (at a given point on a fault). For all sequences the parameters of the best-fit Brownian passage-time, lognormal, and Weibull distributions are given as well as the goodness-of-fit estimators.

| Sequence | Data or fit | Data or fit parameters[†] | | Goodness-of-fit estimators | | | |
|---|---|---|---|---|---|---|---|
| | | | | $D_{KS}$ | $Q_{KS}$ | Cdf RMSE | Weibull plot RMSE |
| xhr2 10 minute 51 events | Sample distribution | $\mu = 2.4$ mm | $C_V = 0.56$ | Not applicable | | | |
| | Brownian passage-time | $\mu = 2.4\pm0.2$ mm | $C_V = 0.72\pm0.08$ | 0.18 | 0.07 | 0.09 | 0.3 |
| | Lognormal | $\bar{y} = 0.71\pm0.09$[*] | $\sigma_y = 0.65\pm0.06$[*] ($C_V = 0.72\pm0.09$) | 0.14 | 0.2 | 0.07 | 0.2 |
| | Weibull | $\tau = 2.7\pm0.2$ mm | $\beta = 1.9\pm0.2$ ($C_V = 0.55\pm0.05$) | 0.09 | 0.8 | 0.04 | 0.09 |
| cwn1 10 minute 45 events | Sample distribution | $\mu = 3.4$ mm | $C_V = 0.48$ | Not applicable | | | |
| | Brownian passage-time | $\mu = 3.4\pm0.3$ mm | $C_V = 0.60\pm0.07$ | 0.15 | 0.2 | 0.07 | 0.4 |
| | Lognormal | $\bar{y} = 1.10\pm0.08$[*] | $\sigma_y = 0.55\pm0.06$[*] ($C_V = 0.59\pm0.07$) | 0.13 | 0.4 | 0.06 | 0.3 |
| | Weibull | $\tau = 3.9\pm0.3$ mm | $\beta = 2.2\pm0.2$ ($C_V = 0.47\pm0.05$) | 0.07 | 0.97 | 0.03 | 0.08 |
| xhr2 daily 76 events | Sample distribution | $\mu = 2.6$ mm | $C_V = 0.58$ | Not applicable | | | |
| | Brownian passage-time | $\mu = 2.6\pm0.2$ mm | $C_V = 0.83\pm0.08$ | 0.19 | 0.007 | 0.10 | 0.2 |
| | Lognormal | $\bar{y} = 0.73\pm0.08^{*}$ | $\sigma_y = 0.73\pm0.06^{*}$ ($C_V = 0.83\pm0.09$) | 0.15 | 0.07 | 0.07 | 0.19 |
| | Weibull | $\tau = 2.9\pm0.2$ mm | $\beta = 1.78\pm0.16$ ($C_V = 0.58\pm0.05$) | 0.08 | 0.7 | 0.04 | 0.09 |
| cwn1 daily 104 events | Sample distribution | $\mu = 3.2$ mm | $C_V = 0.52$ | Not applicable | | | |
| | Brownian passage-time | $\mu = 3.2\pm0.3$ mm | $C_V = 0.81\pm0.07$ | 0.18 | 0.002 | 0.10 | 0.3 |
| | Lognormal | $\bar{y} = 0.98\pm0.07^{*}$ | $\sigma_y = 0.70\pm0.05^{*}$ ($C_V = 0.80\pm0.07$) | 0.13 | 0.05 | 0.07 | 0.3 |
| | Weibull | $\tau = 3.63\pm0.19$ mm | $\beta = 2.00\pm0.16$ ($C_V = 0.52\pm0.04$) | 0.10 | 0.3 | 0.04 | 0.12 |

[†] The error bars are 95% confidence limits.
[*] Units of data in mm

Table 2. Slider-block model: The recurrent slip amplitude statistics at a given point of the model. For all sequences the parameters of the best-fit Brownian passage-time, lognormal, and Weibull distributions are given as well as the goodness-of-fit estimators. Units of slip amplitude are non-dimensional, introduced by Eqs.(17).

| Sequence | Data or fit | Data or fit parameters[†] | | Goodness-of-fit estimators | | | |
|---|---|---|---|---|---|---|---|
| | | | | $D_{KS}$ | $Q_{KS}$ | Cdf RMSE | Weibull plot RMSE |
| Strongest block 734 events | Sample distribution | $\mu = 0.128$ | $C_V = 0.64$ | Not applicable | | | |
| | Brownian passage-time | $\mu = 0.128 \pm 0.005$ | $C_V = 1.03 \pm 0.03$ | 0.16 | $1.2 \cdot 10^{-16}$ | 0.10 | 0.6 |
| | Lognormal | $\overline{y} = -2.31 \pm 0.03$ | $\sigma_y = 0.80 \pm 0.02$ ($C_V = 0.95 \pm 0.03$) | 0.07 | 0.0007 | 0.04 | 0.3 |
| | Weibull | $\tau = 0.142 \pm 0.003$ | $\beta = 1.61 \pm 0.05$ ($C_V = 0.638 \pm 0.017$) | 0.02 | 0.87 | 0.009 | 0.04 |
| Weakest block 730 events | Sample distribution | $\mu = 0.127$ | $C_V = 0.63$ | Not applicable | | | |
| | Brownian passage-time | $\mu = 0.127 \pm 0.005$ | $C_V = 1.03 \pm 0.03$ | 0.16 | $9 \cdot 10^{-18}$ | 0.10 | 0.6 |
| | Lognormal | $\overline{y} = -2.30 \pm 0.03$ | $\sigma_y = 0.80 \pm 0.02$ ($C_V = 0.94 \pm 0.03$) | 0.08 | $8 \cdot 10^{-5}$ | 0.04 | 0.3 |

| | Weibull | $\tau = 0.142\pm0.003$ | $\beta = 1.63\pm0.05$ <br><br> ($C_V = 0.629\pm0.017$) | 0.019 | 0.96 | 0.006 | 0.04 |
|---|---|---|---|---|---|---|---|
| Medium block <br><br> 733 events | Sample distribution | $\mu = 0.127$ | $C_V = 0.62$ | Not applicable | | | |
| | Brownian passage-time | $\mu = 0.127\pm0.005$ | $C_V = 1.01\pm0.03$ | 0.17 | $1.2\cdot10^{-19}$ | 0.10 | 0.6 |
| | Lognormal | $\bar{y} = -2.30\pm0.03$ | $\sigma_y = 0.79\pm0.02$ <br><br> ($C_V = 0.93\pm0.03$) | 0.09 | $1.7\cdot10^{-5}$ | 0.05 | 0.3 |
| | Weibull | $\tau = 0.142\pm0.003$ | $\beta = 1.65\pm0.05$ <br><br> ($C_V = 0.621\pm0.016$) | 0.03 | 0.7 | 0.011 | 0.05 |

[†] The error bars are 95% confidence limits.

Table 3. Slider-block model: The cumulative distribution of recurrent energies for the system-wide event sequence (at a given fault). The parameters of the best-fit Brownian passage-time, lognormal, and Weibull distributions are given as well as the goodness-of-fit estimators. Units of energy are non-dimensional, introduced by Eqs.(17).

| Sequence | Data or fit | Data or fit parameters[†] | | Goodness-of-fit estimators | | | |
|---|---|---|---|---|---|---|---|
| | | | | $D_{KS}$ | $Q_{KS}$ | Cdf RMSE | Weibull plot RMSE |
| Energies dissipated during 715 system-wide events | Sample distribution | $\mu = 94.$ | $C_V = 0.61$ | Not applicable | | | |
| | Brownian passage-time | $\mu = 94.\pm3.$ | $C_V = 0.95\pm0.03$ | 0.14 | $1.4\cdot10^{-13}$ | 0.09 | 0.5 |
| | Lognormal | $\bar{y} = 4.31\pm0.03$ | $\sigma_y = 0.76\pm0.02$ ($C_V = 0.88\pm0.03$) | 0.07 | 0.0009 | 0.04 | 0.3 |
| | Weibull | $\tau = 105.\pm2.$ | $\beta = 1.70\pm0.05$ ($C_V = 0.607\pm0.016$) | 0.02 | 0.88 | 0.009 | 0.05 |

[†] The error bars are 95% confidence limits.

Table 4. Sand-pile model: The cumulative recurrent frequency-size distribution for the sequence of percolating events (at a given point of the model). The parameters of the best-fit Brownian passage-time, lognormal, and Weibull distributions are given as well as the goodness-of-fit estimators.

| Sequence | Data or fit | Data or fit parameters[†] | | Goodness-of-fit estimators | | | |
|---|---|---|---|---|---|---|---|
| | | | | $D_{KS}$ | $Q_{KS}$ | Cdf RMSE | Weibull plot RMSE |
| 2013 left-right-up-down percolating events | Sample distribution | $\mu = 6720.$ | $C_V = 0.143$ | Not applicable | | | |
| | Brownian passage-time | $\mu = 6720.\pm20.$ | $C_V = 0.152\pm0.002$ | 0.05 | 0.00018 | 0.02 | 0.2 |
| | Lognormal | $\bar{y} = 8.80\pm0.003$ | $\sigma_y = 0.151\pm0.002$ ($C_V = 0.152\pm0.002$) | 0.05 | $1.6\cdot10^{-5}$ | 0.03 | 0.19 |
| | Weibull | $\tau = 7130.\pm20.$ | $\beta = 8.09\pm0.14$ ($C_V = 0.147\pm0.002$) | 0.02 | 0.3 | 0.011 | 0.04 |

[†] The error bars are 95% confidence limits.

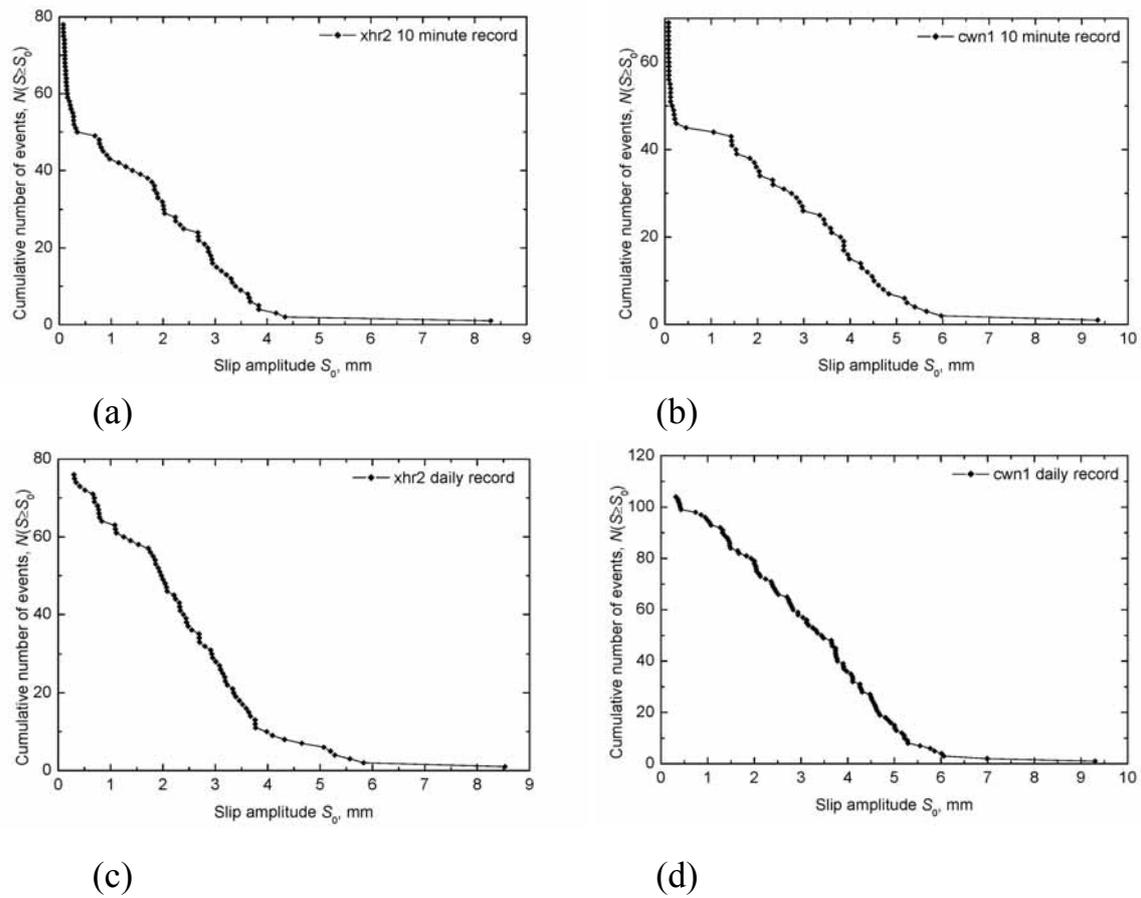

(a)

(b)

(c)

(d)

Figure 1. Creep events (at a given point on a fault): The recurrent cumulative frequency-amplitude distributions of slip amplitudes for the (a) xhr2 10 minute record, (b) cwn1 10 minute record, (c) xhr2 daily record, and (d) cwn1 daily record. The plots are constructed in the traditional Gutenberg-Richter style. The cumulative distribution functions are integrated from large to small amplitudes and are not normalized.

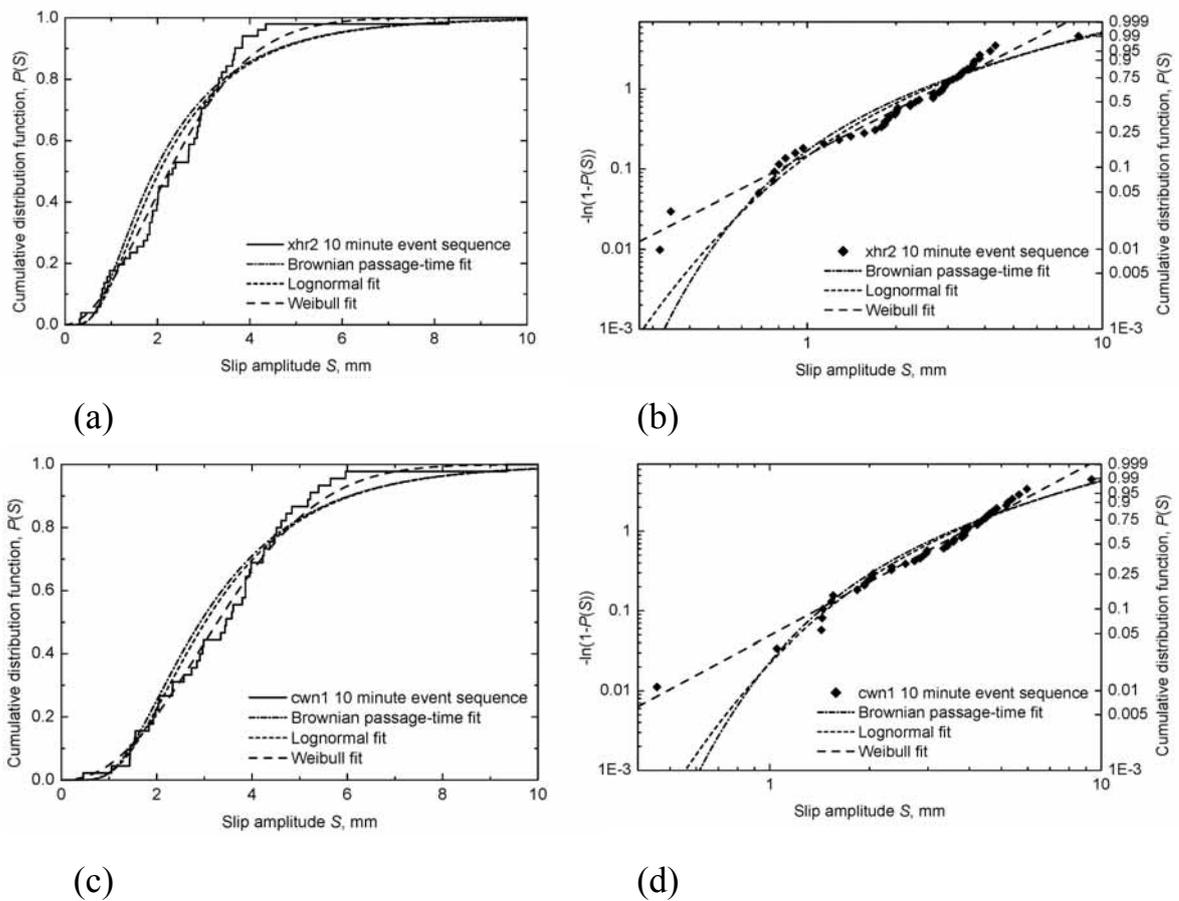

(a)

(b)

(c)

(d)

Figure 2. Creep events (at a given point on a fault): The recurrent cumulative frequency-amplitude distributions for the sequences of 51 slip amplitudes of the xhr2 10 minute record, (a) and (b), and of 45 slip amplitudes of the cwn1 10 minute record, (c) and (d). In (a) and (c) the cumulative distribution functions of recurrent slip amplitudes are given as solid lines. The corresponding Weibull plots are given in (b) and (d) as diamonds. In all cases the data are compared with the best-fit Brownian passage-time distributions (dash-dot lines), the best-fit lognormal distributions (short-dash lines), and the best-fit Weibull distributions (long-dash lines).

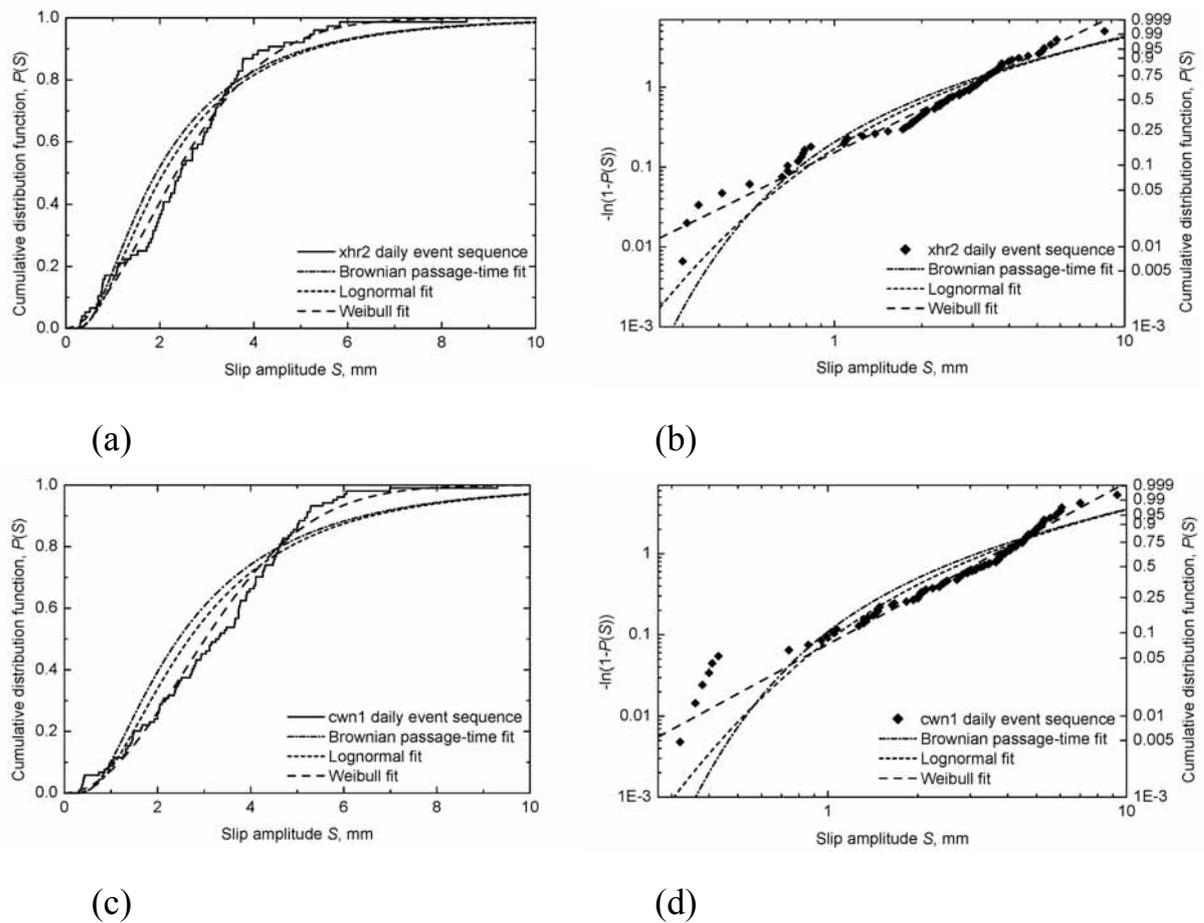

(a)

(b)

(c)

(d)

Figure 3. Creep events (at a given point on a fault): The recurrent cumulative frequency-amplitude distributions for the sequences of 76 slip amplitudes of the xhr2 daily record, (a) and (b), and of 104 slip amplitudes of the cwn1 daily record, (c) and (d). In (a) and (c) the cumulative distribution functions of recurrent slip amplitudes are given as solid lines. The corresponding Weibull plots are given in (b) and (d) as diamonds. In all cases the data are compared with the best-fit Brownian passage-time distributions (dash-dot lines), the best-fit lognormal distributions (short-dash lines), and the best-fit Weibull distributions (long-dash lines).

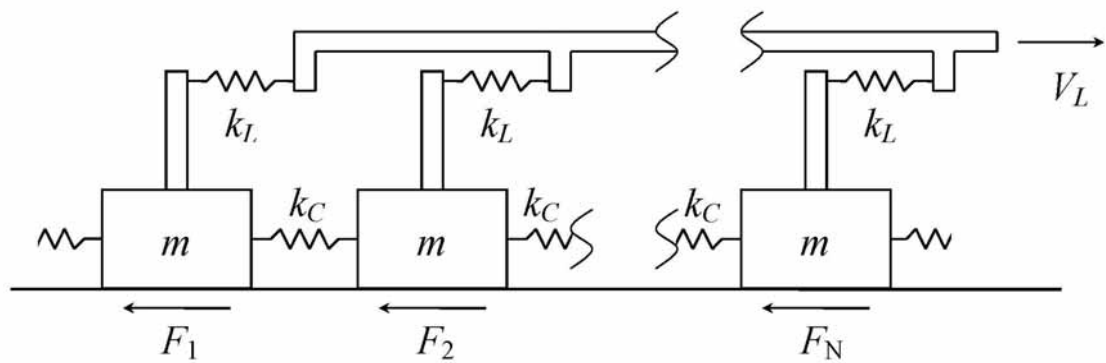

Figure 4. Illustration of one-dimensional slider-block model. A linear array of $N = 500$ blocks of mass $m$ is pulled along a surface at a constant velocity $V_L$ by a loader plate. The loader plate is connected to each block with a loader spring with spring constant $k_L$ and adjacent blocks are connected by springs with spring constant $k_C$. The frictional resisting forces are $F_1, F_2, \ldots, F_N$.

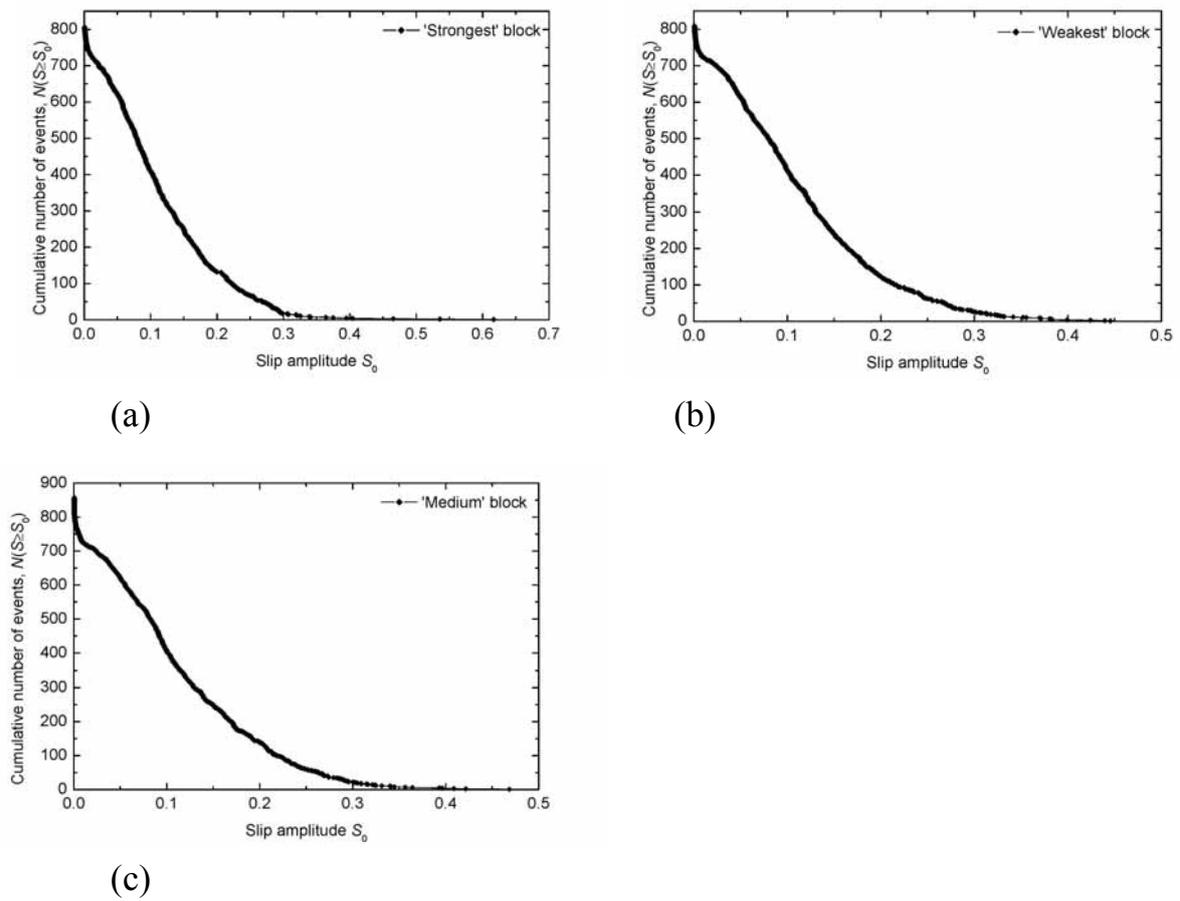

(a)                                 (b)

(c)

Figure 5. Slider-block model: The recurrent cumulative frequency-amplitude distributions for event sequences of the (a) strongest block, (b) weakest block, and (c) medium block (at a given point of the model). The plots are constructed in the traditional Gutenberg-Richter style. The cumulative distribution functions are integrated from large to small amplitudes and are not normalized.

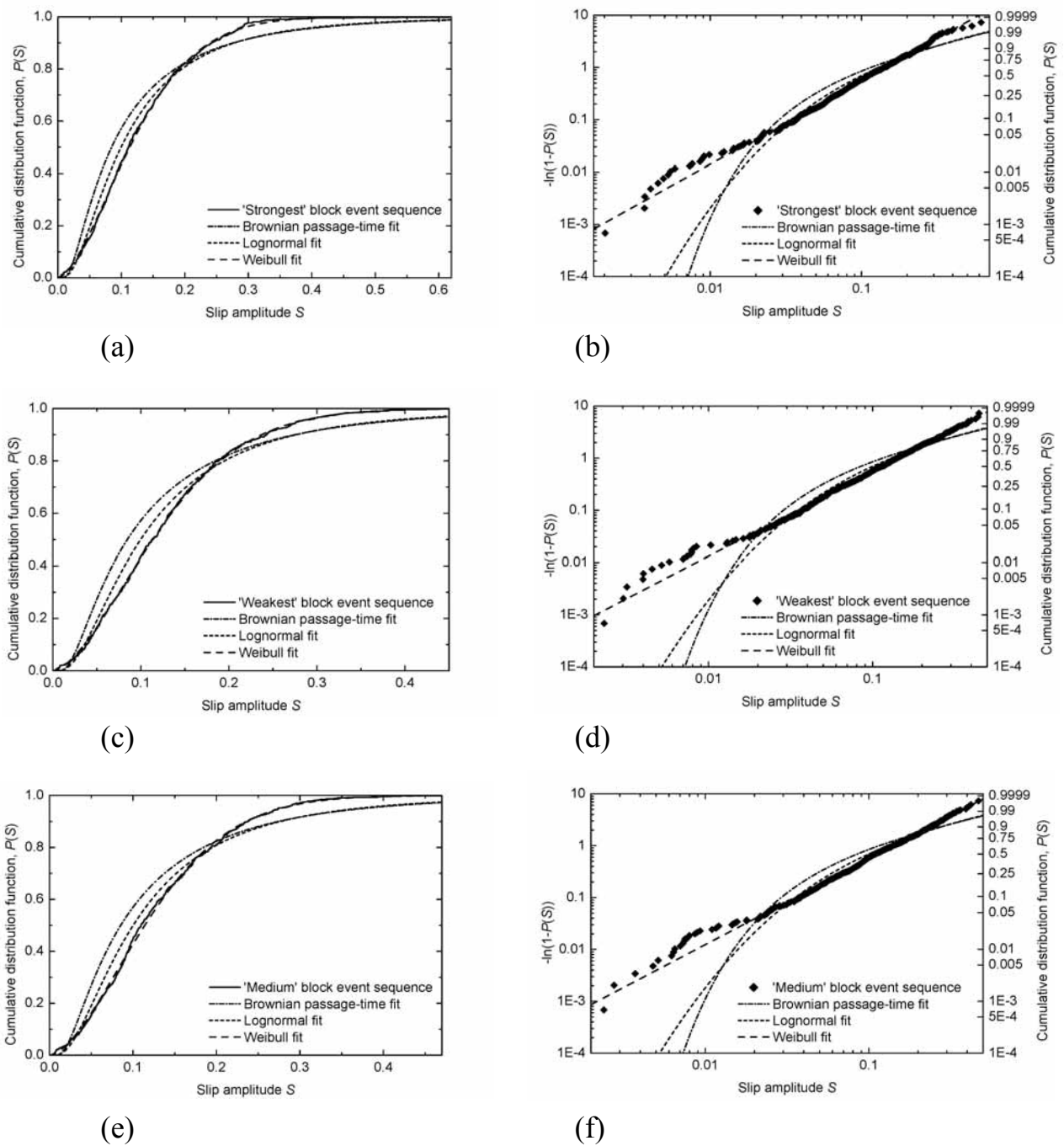

(a)

(b)

(c)

(d)

(e)

(f)

Figure 6. Slider-block model: The recurrent cumulative frequency-amplitude distributions for the sequences of 734 slip amplitudes of the strongest block, (a) and (b), of 730 slip amplitudes of the weakest block, (c) and (d), and of 733 slip amplitudes of the medium block (at a given point of the model). In (a), (c), and (e) the cumulative distribution functions of recurrent slip amplitudes are given as solid lines. The corresponding Weibull plots are given in (b), (d), and (f) as diamonds. In all cases the data are compared with the best-fit Brownian passage-time distributions (dash-dot lines), the best-fit lognormal distributions (short-dash lines), and the best-fit Weibull distributions (long-dash lines).

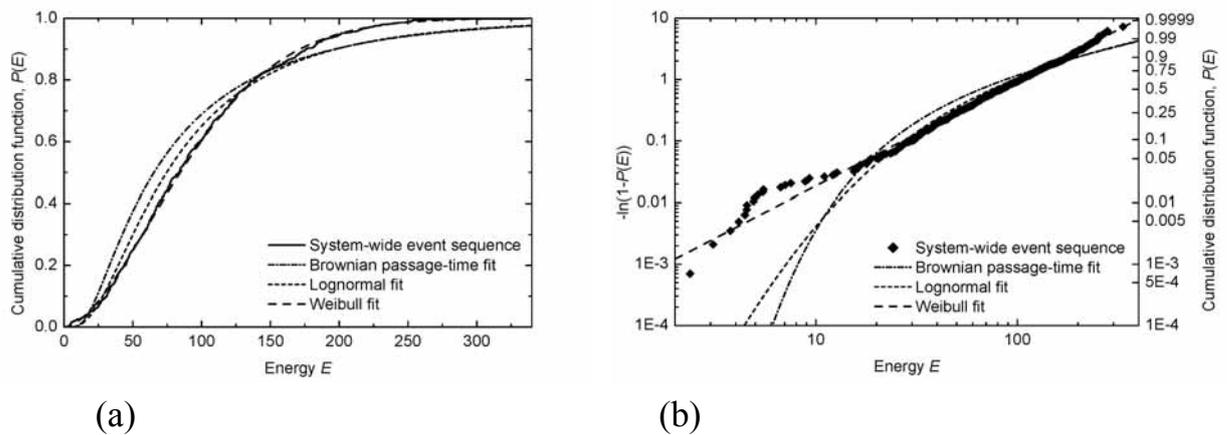

| (a) | (b) |

Figure 7. Slider-block model: The recurrent cumulative frequency-energy distribution for the sequence of 715 system-wide events (at a given fault). In (a) the cumulative distribution function of recurrent energies is given as a solid line. The corresponding Weibull plot is given in (b) as diamonds. In both cases the data are compared with the best-fit Brownian passage-time distribution (dash-dot lines), the best-fit lognormal distribution (short-dash lines), and the best-fit Weibull distribution (long-dash lines).

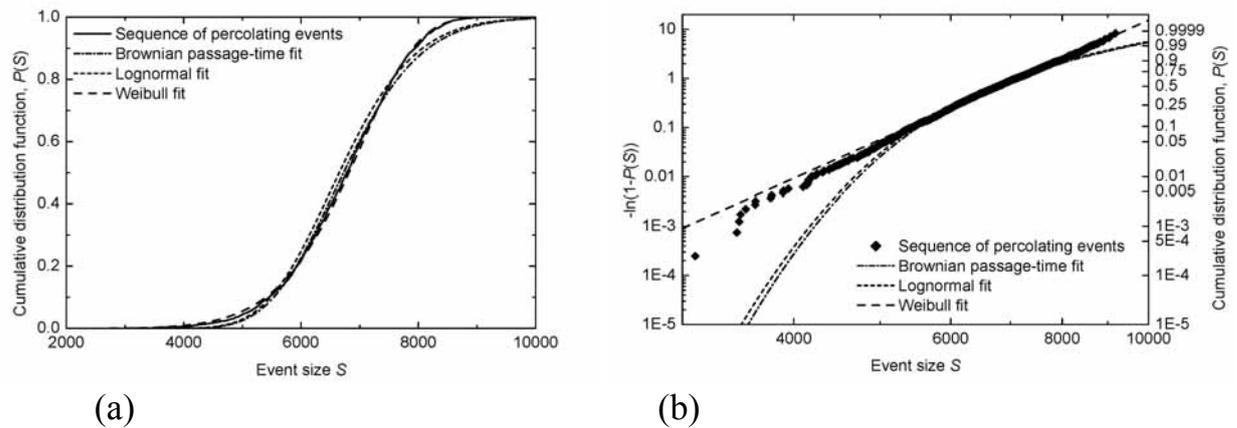

(a)     (b)

Figure 8. Sand-pile model: The recurrent cumulative frequency-size distribution for the sequence of 2013 percolating events (at a given point of the model). In (a) the cumulative distribution function of recurrent event sizes is given as a solid line. The corresponding Weibull plot is given in (b) as diamonds. In both cases the data are compared with the best-fit Brownian passage-time distribution (dash-dot lines), the best-fit lognormal distribution (short-dash lines), and the best-fit Weibull distribution (long-dash lines).